\documentclass[11pt,a4paper,english,nofootinbib,,superscriptaddress]{revtex4}
\usepackage{lmodern}

\usepackage[T1]{fontenc}
\usepackage[latin9]{inputenc}
\usepackage{babel}
\usepackage{color}
\usepackage{amsmath}
\usepackage{graphicx}
\usepackage{amssymb}
\usepackage{esint}
\usepackage[unicode=true, pdfusetitle,
 bookmarks=true,bookmarksnumbered=false,bookmarksopen=false,
 breaklinks=false,pdfborder={0 0 1},backref=false,colorlinks=false]
 {hyperref}
\usepackage{amsmath}
\usepackage{amssymb}
\def\b{\begin{equation}} \def\e{\end{equation}}
\def\bd{\begin{displaystyle}} \def\ed{\end{displaystyle}}
\def\ba{\begin{array}} \def\ea{\end{array}}

\def\bee{\begin{enumerate}}
\def\eee{\end{enumerate}}

\def\N{\mbox{I\hspace{-.15em}N}}

\def\1{\mbox{I\hspace{-.15em}1}}

\def\R{{\rm I\hspace{-.15em}R}}

\def\C{\hspace{3pt}{\rm l\hspace{-.47em}C}}

\def\b{\begin{equation}}
\def\e{\end{equation}}
\def\bee{\begin{enumerate}}
\def\eee{\end{enumerate}}

\makeatletter
\usepackage{latexsym}\usepackage{bm}

\makeatother

\begin{document}
\title{Conformally Covariant Vector-Spinor Field in de Sitter Space}

\author{N. Fatahi}
\email{fatahi@iauksh.ac.ir } \affiliation{Department of Physics,
Science and Research branch, Islamic Azad University, Tehran,
Iran}
\author{M.V. Takook}
\email{takook@razi.ac.ir} \affiliation{Department of Physics,
Science and Research branch, Islamic Azad University, Tehran,
Iran}
\author{M.R. Tanhayi}
\email{m_tanhayi@iauctb.ac.ir} \affiliation{Department of Physics,
Islamic Azad University, Central Tehran Branch, Tehran, Iran}
\date{\today}

\begin{abstract}

\noindent \hspace{0.35cm}

In this paper, we study conformally invariant field equations for
vector-spinor (spin-$\frac{3}{2}$) field in de Sitter space-time.
The solutions are also obtained in terms of the de Sitter-Dirac
plane waves. The related two-point functions are calculated in
both de Sitter ambient space formalism and intrinsic coordinate.
In order to study the conformal invariance, Dirac's six-cone
formalism is utilized in which the field equations are expressed
in a manifestly conformal way in $4+2$ dimensional conformal space
and then followed by the projection to de Sitter space.

\end{abstract}

\maketitle
\section{Introduction}

Recent astrophysical data indicate that our
universe is undergoing an accelerated expansion \cite{Riess} and at the first approximation the background space-time can be well described by the de Sitter (dS) space. The dS universe is a maximally symmetric curved space with the same degrees of symmetry as the Minkowskian space. It also plays an important role in the inflation scenario where the cosmic dynamics is assumed to be dominated by a term that acting like a cosmological constant \cite{linde}. Thus, constructing a formulation of dS quantum field theory seems to be important which has been studied by many authors \cite{qft, qft1, qft2, qft3, entropy,ta1403}. The high level of symmetry in this curved space-time can be a guideline for constructing the quantum fields and the underlining symmetry of the background may be well established by group theory. Actually revealing the group theoretical and geometrical structures underlying the theory is of great importance. We use this idea to study the higher-spin free field in de Sitter space. Basically studying the free higher-spin fields may shed some light to our better understanding of the interactions in higher-spin field theories. As one knows, the construction of a consistent interacting theory of higher-spin fields is one of the oldest long standing problems in theoretical physics of particles and fields.

The higher-spin field equations have been considered in \cite{wigner1}. It was proved that a given field of rest mass $m$ and spin $s \geq 1$ can be represented by a completely symmetric multi-spinor \cite{wigner3}. Manifestly covariant massless fields for half-integer spin in Anti-de Sitter background have been studied in \cite{fronsdal12}. Massive spin-$\frac{3}{2}$ field in dS space has been considered in \cite{baba} and the extension to the massless fields is done in \cite{azizi}. Here, within the language of the group theory in de Sitter space, we study the conformal invariance of vector-spinor field by finding the field equations, solutions and two-point function.

As it is known, the massless fields which live on the light-cone, are conformally invariant \cite{Wald}, generally, all massless gauge fields in four dimensional constant curvature spaces are conformally invariant at the level of degrees of freedom. This indicates that the respective representations of (A)dS or Poincar\'e group can be extended to the representations of the conformal group \cite{ta1403}. The details for higher-spin gauge fields can be found in \cite{new,new2}. For our purpose, in dS space-time the conformal invariance may be understood by the concept of light-cone propagators. Noting that in dS space, mass is not a well-defined invariant parameter for a set of observers transforming under the dS group $SO(1,4)$,  the masslessness is used in reference to the conformal invariance (propagation on the dS light-cone). From the group theoretical point of view, the massive representations refer to those that coincide to the massive representation of the Poincar\'e group at the zero curvature limit (appendix A). Our motivation to study the spin-vector conformal invariance field equation in dS space is that in the supergravity model this field is the partner of the gravitational field and it must propagate on the dS light-cone.

In flat space, according to the group representation theory and Wigner's interpretation of the elementary systems, the field operators of any kind of spin transform according to the unitary irreducible representation (UIR) of the Poincar\'e group (the kinematical group of Minkowski space), and their corresponding field equations are well established by the Casimir operators. Being interested in symmetry of space, the isometry group of dS space is $SO_0(1,4)$ which may be viewed as a deformation of the proper orthochronous Poincar\'e group. There are two Casimir operators in dS group and it has been shown that the massive scalar, vector and spin-$2$ fields can be associated with the UIRs of dS group \cite{qft2,re,chli,ta1,anilto}. The massless fields can be associated with an indecomposable representation of dS group \cite{gagarota}. The covariant quantum field theory for massive and massless conformally coupled scalar field in dS space have been studied in \cite{bgm,qft2} and also for the massless minimally coupled scalar field in \cite{grt}. In Ref.s \cite{new3,new4}, one can find the conformally covariant quantization of the gauge field in dS space. 

The main goal of this paper is to study conformal invariance of spin-$3/2$ field in de Sitter space, we use group theoretical approach and hence the de Sitter invariance becomes manifest. The paper is organized in seven sections:
The conformally invariant spin-$\frac{3}{2}$ field equations are obtained in section II and III. The solutions are found in section IV, as it will be shown, they can be written in terms of the de Sitter-Dirac plane waves. In section V and VI, we find the two-point function in both embedding space namely $4+1$ ambient space of dS and also in 4 dimensional intrinsic space. We discuss the results in the conclusion part. Finally, some useful relations are presented in the appendixes.

\setcounter{equation}{0}

\section{ massless spinor field equations in de Sitter space}

The dS metric is the unique solution of Einstein's equation in vacuum with
positive cosmological constant $\Lambda = 3 H^2$, in which dS space may be visualized as the hyperboloid embedded in a five-dimensional Minkowski space
\b X_{H}=\left\{x\in
\R^{5};x^{2}=\eta_{\alpha\beta}x^{\alpha}x^{\beta}=-H^{-2}\right\},\;\;
 \;\e
$H$ is the Hubble constant, hereafter for the sake of simplicity, we set $H=1$. The de Sitter
metric is
 \b ds^{2}=\eta_{\alpha\beta}dx^{\alpha}dx^{\beta}\mid_{x^{2}=-1}=g_{\mu\nu}^{dS}dX^{\mu}dX^{\nu},\;\;\e
where $\eta_{\alpha\beta}=diag (1,-1,-1,-1,-1)$, $\alpha,\beta=0,1,2,3,4,$ and $\mu,\nu=0,1,2,3$. We use $x^\alpha$ for ambient space formalism whereas $X$ stand for de Sitter intrinsic coordinates. For
simplicity the dot product is shown as $x^\alpha \partial_\alpha\equiv x\cdot
\partial$. We define the transverse derivative in de Sitter
space as
$\partial^T_\alpha\equiv\theta_{\alpha\beta}\partial^\beta=\partial_\alpha+H^2x_\alpha
x\cdot\partial$, where
$\theta_{\alpha\beta}=\eta_{\alpha\beta}+H^2x_\alpha x_\beta$ is
the projection operator and note that
$x\cdot \partial^T=0$. Working in embedding space has two advantages, first it is close to the group theoretical language and second the equations are obtained in an easer way than they might be found in de Sitter intrinsic space.

There are two Casimir operators for dS
group, these operators commute with all the action of the group
generators and thus they are constant on each representation. In this section we briefly recall the notations of the Casimir operator and more details can be found in \cite{bida, dixmier, sep, re}. The second and forth order Casimir operators are:
 \begin{eqnarray}
Q^{(1)}=-\frac{1}{2}L^{\alpha\beta}L_{\alpha\beta},\hspace{1cm}
Q^{(2)}=-W_{\alpha}W^{\alpha},
\end{eqnarray}
where
$W_{\alpha}=\frac{1}{8}\epsilon_{\alpha\beta\gamma\delta\eta}L^{\beta\gamma}L^{\delta\eta}$
  and $\epsilon_{\alpha\beta\gamma\delta\eta}$ is the
antisymmetric tensor in the ambient space notation with
$\epsilon_{01234}=1$. The generator of de Sitter group is defined by
$$L_{\alpha\beta}=M_{\alpha\beta}+S_{\alpha\beta},$$ where the
"orbital" part is
 \b M_{\alpha\beta}=-i(x_{\alpha}\partial_{\beta}-x_{\beta}\partial_{\alpha})=-i(x_{\alpha}\partial^T_{\beta}
  -x_{\beta}\partial^T_{\alpha}),\e
and the "spinoral" part $S_{\alpha\beta}$ which acts on the spinor field ($s=1/2$) is \cite{bida} \b
S_{\alpha\beta}=-\frac{i}{4}[\gamma_{\alpha},\gamma_{\beta}].\e
In this case the five $4\times 4$ matrices of $\gamma^\alpha$ are
the generators of the Clifford algebra which are constructed
as \b\label{clifford}
\gamma^{\alpha}\gamma^{\beta}+\gamma^{\beta}\gamma^{\alpha}=2\eta^{\alpha\beta}\;\;\;,\;\;
\;\gamma^{\alpha\dag}=\gamma^{0}\gamma^{\alpha}\gamma^{0}.\e

Based on spectrum of the possible values of Casimir eigenvalues, the UIRs of de Sitter group can be classified as the principal, complementary and discrete series (Appendix
A). In the principal series which belongs to the massive
representation of dS space and tends to the massive representation of
Poincar\'e group at the zero curvature limit, the eigenvalues of the
Casimir operators can be written as \cite{dixmier}
\begin{eqnarray}
<Q^{(1)}_s>=\Big(\frac{9}{4}+\nu^{2}-s(s+1)\Big),\nonumber\\
<Q^{(2)}_s>=\Big((\frac{1}{4}+\nu^{2})s(s+1)\Big),\nonumber
\end{eqnarray}
where $s$ stands for the spin, $\nu$ is a real positive
parameter. The second order field equations can be written as\footnote{Note that in writing the field equations we use only the second order Casimir operator, because the forth order one leads to higher derivative equations  \b Q^{(2)}=-W_{\alpha}W^{\alpha}=\frac{1}{64}\epsilon_{\alpha\beta\gamma\delta\eta}\epsilon^{\alpha\beta\gamma\delta\eta}[M^{\beta\gamma}M^{\delta\eta}M_{\beta\gamma}M_{\delta\eta}+M^{\beta\gamma}M^{\delta\eta}M_{\beta\gamma}S_{\delta\eta}\nonumber \e
$$ +M^{\beta\gamma}M^{\delta\eta}S_{\beta\gamma}M_{\delta\eta}+M^{\beta\gamma}M^{\delta\eta}S_{\beta\gamma}S_{\delta\eta}+...]$$}
\b \label{field equation}(Q^{(1)}_s-<Q^{(1)}_s>)\psi=0.\e

For example for $s=1/2$, one has
\b \label{equation}(Q^{(1)}_{\frac{1}{2}}-\frac{3}{2})\psi=\nu^2\psi,\e
 $\psi$ stands for a spinor field with arbitrary degree of homogeneity: $x\cdot \partial \psi=\sigma \psi$.
 The second order Casimir operator for spin-$\frac{1}{2}$ is given by \b
Q^{(1)}_{\frac{1}{2}}=-\frac{1}{2}M_{\alpha\beta}M^{\alpha\beta}-\frac{1}{2}S_{\alpha\beta}S^{\alpha\beta}-S_{\alpha\beta}M^{\alpha\beta},\e
where one can show \b
\frac{1}{2}S_{\alpha\beta}S^{\alpha\beta}=\frac{5}{2},\,\,\,\,\;
\;S_{\alpha\beta}M^{\alpha\beta}=-\frac{i}{2}\gamma_{\alpha}\gamma_{\beta}M^{\alpha\beta}=-\not
x \not
\partial^T,\e
note that $ \gamma_\alpha x^\alpha\equiv \not x$. After making use of above relations, equation (\ref{equation}) can be written in
terms of the scalar Casimir operator as follows
\b\label{12} (Q_{0}+\not x \not
\partial^T-4)\psi=\nu^2\psi,\e
where $Q^{(1)}_{0}(\equiv Q_{0})=-\frac{1}{2}M_{\alpha\beta}M^{\alpha\beta}$ is the spin-less Casimir operator. If one defines the de Sitter-Dirac operator ${\cal D}$ as \b \label{dual12}
{\cal
D}\equiv-\frac{i}{2}\gamma_{\alpha}\gamma_{\beta}M^{\alpha\beta}+2=-\not
x\not\partial^T+2 ,\e  then (\ref{12}) can be written as follows
\b\label{dirac equation}(i{\cal
D}-\nu)(i{\cal D}+\nu)\psi(x)=0\;.\e
This relation is similar to the standard relation of the spinor field in flat space, $
(i\not\partial+m)(i\not\partial-m)\psi(x)=0$. The first order field equation for a field of spin-$\frac{1}{2}$ in dS space becomes
\b(i{\cal D}+\nu)\psi(x)=0,\,\,\,\,\mbox{ where}\,\,\, \nu\neq 0 \in  \R.\e

The massless case in de Sitter space belongs to the discrete series and the eigenvalue of the second order Casimir operator is given by
\b <Q_s^{(1)}>=2(1-s^2)\1.\e
Plugging this value for $s=\frac{1}{2}$ in (\ref{field equation}) and in terms of the ${\cal D}$ operator, one obtains
 \b \label{conformal} i{\cal D}\psi(x)=0,\,\,\, \mbox{or equivalently}\,\,\,\, (Q_0-2)\psi=0. \e
Similarly for spin-$\frac{3}{2}$ field equation in dS space, two types of UIRs of dS group are
characterized, the principal and discrete series
\begin{itemize}
\item[i)] The unitary irreducible representations
$U_{\frac{3}{2},\nu}$ of the principal series,$$
<Q_\frac{3}{2}^{(1)}>=(\nu^2-\frac{3}{2}),\;\;\;\;\nu \in  \R
\:\:\: \nu>\frac{3}{2},$$ note that $U_{\frac{3}{2},\nu}$ and
$U_{\frac{3}{2},-\nu}$ are equivalent this kind of representation belong to the massive case.
\item[ ii)]  The unitary
irreducible representations $\Pi^{\pm}_{\frac{3}{2},\frac{3}{2}}$ of the
discrete series,  \b <Q_\frac{3}{2}^{(1)}>=-\frac{5}{2}, \e
 the sign ${\pm}$ stands for the helicity.
\end{itemize}
The massless spin-$\frac{3}{2}$ field in dS space becomes \cite{baba, azizi}
\begin{equation} \label{dirac1}
\left(Q_\frac{3}{2}^{(1)}+\frac{5}{2}\right)\Psi_{\alpha}(x)=0,
\end{equation}
where
\begin{align}\label{25}
Q^{(1)}_{\frac{3}{2}}\Psi_{\alpha}(x)=&\left(-\frac{1}{2}M_{\alpha
\beta}M^{\alpha
      \beta}+\frac{i}{2} \gamma_{\alpha}\gamma_{\beta}
       M^{\alpha \beta}-\frac{11}{2}\right)\Psi_{\alpha}(x)\nonumber\\
       &-2\partial_\alpha x\cdot\Psi(x)+2x_\alpha\partial\cdot\Psi(x)+\gamma_\alpha
       \gamma\cdot\Psi(x).
\end{align}
Similar to the massless vector-spinor fields equations in Minkowski space, the solutions of this filed equation possess a singularity due to the divergencelessness condition $(\partial^T\cdot\psi=0)$. Then the gauge invariant field equation is \cite{azizi} \b
\label{relation}(Q^{(1)}_{\frac{3}{2}}+\frac{5}{2})
\Psi_\alpha(x)-{\cal D}^{(\frac{3}{2})}_{\alpha} \partial^T
\cdot\Psi(x)=0,\e where ${\cal D}^{(\frac{3}{2})}_{\alpha}\equiv-\partial^T_\alpha-\gamma_\alpha^T\not x.$
This equation is invariant under the following
gauge transformation \b \Psi_\alpha(x)\rightarrow
\Psi'_\alpha(x)=\Psi_\alpha(x)+{\cal
D}^{(\frac{3}{2})}_{\alpha}\zeta,\e where $\zeta$ is an arbitrary
spinor field. In order to fix the gauge, $c$ is introduced and
then one has \b\label{2.28}
\left(Q_\frac{3}{2}^{(1)}+\frac{5}{2}\right)\Psi_\alpha(x)-c{\cal
D}^{(\frac{3}{2})}_{\alpha}
\partial^T \cdot\Psi(x)=0.\e
Similar to the spin-$1/2$ case, one can write the first order field equation as
\begin{equation}
\not x\not\partial^T
\Psi_\alpha(x)-3\Psi_\alpha(x)-x_\alpha\not
x\not\Psi-\partial^T_\alpha\not x \not\Psi=0.
\end{equation}
This equation is
invariant under the gauge transformation
$\Psi_\alpha\rightarrow\Psi'_\alpha=\Psi_\alpha+{\cal
D}^{(\frac{3}{2})}_{\alpha}\zeta$.
There exist another first order field equation \cite{ta1403}:
\begin{equation}\label{2.32}
\not
x\not\partial^T\Psi_\alpha(x)-\Psi_\alpha(x)-x_\alpha\not
x\not\Psi+{\cal D}^{(\frac{3}{2})}_{\alpha}\not
x\not\Psi=0,
\end{equation}
which is
invariant under the gauge transformation
$\Psi_\alpha\rightarrow\Psi'_\alpha=\Psi_\alpha+\partial^T_\alpha\zeta$. It is worth to mention that (\ref{2.32}) appears in the conformal invariant field equation.

\setcounter{equation}{0}
\section{Conformal invariance and Dirac's six-cone formalism}

Massless field equations are expected to be conformally invariant (CI). A trivial example
is the Maxwell's equations where in $1909$ Cunningham and Bateman
showed that these equations are covariant under the larger 15-parameter
conformal group $SO(2,4)$ as well as 10-parameter Poincar\'e group
\cite{stwi}. Fields with spin $s\geq1$ are invariant under the
gauge transformation as well. In $1936$ Dirac used a manifestly
conformally covariant formulation namely, the conformal space notation,
to write down  wave equations in Minkowski space \cite{Dirac}.
The conformal group acts non-linearly on Minkowski coordinates, Dirac used coordinates which the conformal group acts linearly on them. This actually reassembles the conformal space and the theory is
defined on a $d+1$ dimensional hypercone (hereafter named as
Dirac's six-cone) or equivalently, in a $d+2$ dimensional conformal space. Within
this formalism, he obtained scalar, spinor and vector
conformally invariant fields in $d = 4$ flat space-time. This theory
developed in some papers (\cite{Mack} and references therein). The
generalization to dS space was done in \cite{gareta1, ta4} to obtain CI
field equations for scalar, vector and symmetric rank-2 tensor fields. Here, we use this approach to study the spinor fields ($s=\frac{1}{2}\,\,\mbox{and}\,\,\frac{3}{2}$) in de Sitter space. First let us recall this method briefly.

\subsection{Dirac's six-cone}

Basically, the special conformal transformation acts
non-linearly on 4-dimensional coordinates. In conformal space Dirac
proposed the coordinates $u^a$, where $SO(2,4)$
acts linearly on them. Dirac's six-cone is then
defined as a 5-dimensional hyper-surface in $\R^6$ satisfying following constraint
\b
u^{2}=\eta_{ab}u^{a}u^{b}=u_{0}^{2}-\overrightarrow{u}^{2}+u_{5}^{2}=0,\;\; a,b=0,1,...,5,\e
this is obviously invariant under the conformal transformation. A given operator say as $\hat{A}$ is said to be intrinsic if it satisfies
$$
\hat{A}u^{2}\phi=u^{2}\hat{\acute{A}}\phi,$$
where $\phi$ is a function in $\R^{6}$. One should write all the wave equations, subsidiary conditions and etc., in terms of operators that are defined intrinsically on the cone. The following CI system which
is defined on the cone is well established this goal\footnote{In fact, this approach to conformal symmetry leads to the best path to exploit the physical symmetry and it provides a rather simple way to write the conformally invariant field equations. Moreover, it is important to mention that on the cone $u^2=0$, the second-order
Casimir operator of conformal group, ${\cal Q}_2 $, is not a
suitable operator to obtain CI wave equations. Because it is proved on the cone $u^2=0$, it reduces to a constant, consequently, this
operator cannot lead to the wave equations on the cone. The
well-defined operators exist only in exceptional cases. For tensor
fields of degree $ -1,0,1,...$, the intrinsic wave operators are $
\partial^2, (\partial^2)^2, (\partial^2)^3,...$ respectively. This method previously was established for $3 + 2$-de Sitter linear gravity in Ref. \cite{new5}.}
\cite{gareta1} \b\label{4.2}\left\{\begin{array}{l}
  (\partial_{a}\partial^{a})^{n}\Phi=0\;,\\
  \hat{N}_{5}\Phi=(n-2)\Phi\;,\\
\end{array}
 \right.\e
where the powers of d'Alembertian $(\partial_{a}\partial^{a})^{n}$
act intrinsically on fields of conformal degree $(n - 2)$, and
$\Phi$ is a tensor or spinor field of a definite rank
 and symmetry. The conformal-degree operator $\hat N_{5}$ is given by \b \label{hat}\hat N_{5}\equiv u^{a}\partial_{a}.\e
 One can add the following CI conditions to restrict the space of solution
 \begin{itemize}
 \item[i)] transversality: $ u_{a}\Phi^{ab...}=0\;,$
 \item[ii)]  divergencelessness: $ \nabla_{a}\Phi^{ab...}=0\;,$
 \item[iii)] tracelessness: $ \Phi^{a}_{ab...}=0\;,$
 \item[iv)] for tensor-spinor field: $\gamma_a\Phi^{ab...}=0\;.$
 \end{itemize}
In conformal coordinate $u$, the definition of $\nabla$ is given by \cite{il}
 $$ \nabla_a\equiv
u_a\partial_b\partial^b-(2\hat{N_5}+4)\partial_a.$$

The quantities which are evaluated on the cone should be projected to $4+1$
de Sitter space, first, one needs a relation between the coordinates
\b \left\{\begin{array}{l}
x^{\alpha}=(u^{5})^{-1}u^{\alpha}\;,\\
x^{5}=u^{5}\;,\\
\end{array}
\label{coordinates}
 \right.\e
therefore, the intrinsic operators turn to
\cite{gareta1, il}: \begin{eqnarray}\label{lan}
\hat{N}_{5}&=&x_{5}\frac{\partial}{\partial x_{5}},\nonumber \\
(\partial_{a}\partial^{a})^{n}&=&-x_{5}^{-2n}\prod_{j=1}^{n}[Q_{0}+(j+1)(j-2)],
\end{eqnarray}
$$ \nabla_\alpha= - x^{-1}_{5} \left[x_\alpha [Q_0-\hat{N_5}(\hat{N_5}-1)]+2\partial^T_\alpha(\hat{N_5}+1)\right],$$
where $Q_0$ is the Casimir operator in de Sitter space. First, one should write the equations in $u$ coordinates where the conformal invariance is manifest (\ref{4.2}). Then by the help of the above mentioned relations, the obtained equations are related to the embedding space ones. Finally, the desired de Sitter relations can be obtained via the projection. This approach provides a simple way to study the conformal invariance in de Sitter space.

\subsection{conformally invariant spinor field equations}

\textbf{Spinor field $s=\frac{1}{2}$:} The simplest conformally invariant system is
obtained by setting $n=1$ and $\psi\equiv x_{5}\Phi$ in (\ref{4.2}) which after making use of (\ref{lan}) and in language of the Casimir operator of dS group, it turns to
 \b (Q_{0}-2)\psi(x)=0\;.\e
This equation is obviously conformally invariant, and $\psi(x)$
stands for a massless conformally spinor field in dS space (see
Eq.\ref{conformal}). After making use of (\ref{12}), the first
order field equation in this case, becomes as follows
\begin{equation} \left( \not x \not \partial^T-2\right)\psi(x)=0.\end{equation}
Indeed, the field $\psi(x)$, associates with the UIR of dS group
$\Pi^{\pm}_{\frac{1}{2},\frac{1}{2}}$ and propagates on the dS light cone.

\textbf{Vector-spinor $s=\frac{3}{2}$:}  In
this case one should classify the degrees of freedom of vector-spinor
field on the cone in terms of the dS fields, this can be done as below \b\label{kappa}
\Psi_{\alpha}=x_{5}(\Phi_{\alpha}+x_{\alpha}x\cdot\Phi)\;\;,\;\; {\cal K} _{1}=x_{5}\Phi_{5}\;\;,\;\; {\cal K}_{2}=x_{5}x\cdot\Phi,\e
where  ${\cal K}_{1}$ and ${\cal K}_{2}$ are two spinor fields and $\Psi_\alpha$ is a vector-spinor field. Note that $x^\alpha\Psi_\alpha=0$ indicates that $\Psi_\alpha$
lives on dS hyperboloid. To
obtain CI field equations by choosing $n=1$ in (\ref{4.2}), one obtains
\b(Q_0-2)\Phi_\alpha=0,\label{c1}\e
where $\Phi_\alpha $ is a spin-$\frac{3}{2}$ field on the cone.
After doing some tedious but straightforward calculation which partially is
given in appendix C, following CI system of field equations is obtained
  \begin{equation} \left\{
  \begin{array}{l}
  (Q_{0}-2){\cal K}_{1}=0,\\
  (Q_{0}-2){\cal K}_{2}=0,\\
  (Q_{0}-2)\Psi_{\alpha}+2x_{\alpha}{\partial^T}\cdot\Psi+{\partial}_{\alpha}^T{\partial^T}\cdot\Psi=0\label{5.4}
  \end{array}\right.
  \end{equation}
 that indicates: ${\cal K}_1$ and ${\cal K}_2$ are both CI massless spinor fields \cite{gareta1}. By using the transversality condition on the cone $(u_a \Phi^{ab...}=0=u_5\Phi^5+u_\alpha\Phi^\alpha)$ and (\ref{c1}) we obtain (see appendix C)
  \b (Q_{0}-2){\partial^T}.\Psi=0. \label{5.5}\e
However, one can use (\ref{25}) to write the third line of (\ref{5.4}) as
 \b\label{5.7} \Big(Q^{(1)}_{\frac{3}{2}}-\not x\not {\partial^T}+\frac{7}{2}\Big)\Psi_{\alpha}+{\partial}_{\alpha}^T{\partial^T}\cdot\Psi-\gamma_{\alpha}
 \not \Psi=0\;.\e
Using the equations (\ref{dirac1},\ref{2.32}) the above
field equation can be rewritten as  \b \label{5.72}
\Big(Q^{(1)}_{\frac{3}{2}}+\frac{5}{2}\Big)\Psi_{\alpha}-\left(\not
x\not {\partial^T}\Psi_\alpha-\Psi_\alpha-x_\alpha \not x \not
\Psi-{\partial}_{\alpha}^T{\partial^T}\cdot\Psi+\gamma_{\alpha}^T
 \not \Psi\right)=0\;.\e
As previously mentioned, the fields should be projected to the de
Sitter space, the transverse projection implies the transversality
of fields, $x\cdot \Psi=0$, so that from the homogeneity
condition, one obtains $ x^\alpha x\cdot\partial \Psi_\alpha=0$.
These two conditions impose the following constrains on the
projected fields: $x\cdot \Phi=0=\Phi^5$, and consequently $
\Psi_\alpha=\Phi_\alpha.$ At the appendix C, it is shown that the CI divergencelessness
condition on the cone, namely $ \nabla_a \Phi^{a}=0 $, results in
${\partial^T}_\alpha\Psi^\alpha=0={\partial^T}_\alpha\Phi^\alpha
$, which indicates the divergenceless fields are only mapped from
the cone on dS hyperboloid.

For simplicity and irreducibility of vector-spinor field representation, the CI condition $\gamma^a \Phi_a=0$ on the cone is imposed, this leads to  $\gamma^\beta\Psi_\beta=\not\Psi=0$, which is the conformally invariant condition on the de Sitter hyperboloid. Imposing this condition and irreducibility (see ( \ref{dirac1})), one receives the following first and second order CI field equations
 \begin{equation}
\label{4.212}\left(\not x\not\partial^T-1\right)\Psi_\alpha(x)=0,\,\,\,\,\,\,\mbox{and}\,\,\,\,\,\, \left(Q^{(1)}_{\frac{3}{2}}+\frac{5}{2}\right)\Psi_{\alpha}=0.
\end{equation}
In this case $\Psi_{\alpha}$ associates with the UIR of dS group, namely
$\Pi^{\pm}_{\frac{3}{2},\frac{3}{2}}$, and note that it propagates on the dS light cone. In the following sections, we find the solution and also obtain the two-point function of this vector-spinor field.

\setcounter{equation}{0}
 \section{Solutions of the conformally invariant wave equations}

A general solution of the field equation (\ref{4.212}) can be written in terms of three spinor fields $\psi_1,\psi_2$ and $\psi_3 $ as follows
 \b\label{6.1} \Psi_\alpha (x)= Z^T_\alpha\psi_1+{\cal D}^{(\frac{3}{2})}_{\alpha} \psi_2+\gamma^T_\alpha\psi_3, \e
where $Z^\alpha$ is an arbitrary five-component constant vector field
and $x\cdot Z^T=0.$ Now we should identify the introduced spinor fields $\psi_1,\psi_2$ and $\psi_3 $. If one demands that $ \Psi_\alpha $
satisfies the second order field equation (\ref{4.212}), then
the spinor fields $\psi_1,\psi_2$ and $\psi_3$ obey the following
equations \begin{equation}
\begin{array}{l} \label{rel.s}
(Q_0+\not x\not\partial^T-3)\psi_1=0,\\
2(x\cdot Z)\psi_1+ (Q_0+\not
x\not\partial^T)\psi_2=0,\\
 \left[\not x(Z.x)+\not Z\right]\psi_1+(Q_0+\not
x\not\partial^T)\psi_3=0.
\end{array}
\end{equation}
On the other hand, $ \Psi_\alpha $ must satisfy the first order field equation (\ref{4.212}), therefore one obtains:
 \begin{equation} \label{confirseq}
 \begin{array}{l}
 (\not x \not\partial^T-1)\psi_1=0,\\
\left(\not x-2\right)\psi_2-2 \not
x \psi_3=0,\\
  \not x \not\partial^T\psi_3-\left(4\not x +1\right)\psi_2- \not x x.Z\psi_1=0.
 \end{array}
 \end{equation}
The second line of equation (\ref{confirseq}) result to:
\b \label{psi2psi3} \psi_2=\frac{2}{5}\left(1-2\not x\right)\psi_3,\;\;\;\mbox{or}\,\,\,\, \psi_3=\frac{1}{2}\left(1+2\not x\right)\psi_2,\e
From the equations (\ref{rel.s}) and (\ref{confirseq}), $\psi_1$ satisfies the following first and second order field equations
\b\label{psi1}(Q_0-2)\psi_1=0,\;\;\; (\not x \not\partial^T-1)\psi_1=0,\e
that indicates it can be regarded as a massless conformally
coupled spinor field with homogeneity degree of $\sigma=-1$ and $-2$ \cite{sep}. On the other hand, the homogeneity consideration reveals that the degrees of homogeneity of $\psi_3$ and $\psi_1$ are equal, note that all the sentences in (\ref{6.1}) have the same degrees of homogeneity and also the degrees of homogeneity $Z^T _\alpha $ and $ \gamma^T_\alpha $ are zero.

Now let us multiply $ \not \partial^T $ from the left on  equations (\ref{confirseq}), after doing some calculations, this yields
\begin{eqnarray}
\label{new12}
Q_0\psi_2=\frac{2}{5}(1-2\not x)\psi_2+2\left[ 3x.Z-\not x \not Z^T\right]\psi_1,\\ \label{new22}
Q_0\psi_3=-\frac{2}{5}(2\not x+7)\psi_3+\left[ 3x.Z-\not x \not Z^T\right]\psi_1.
\end{eqnarray}
Inserting these results in equations
(\ref{rel.s}), and after making use of (\ref{psi2psi3}), one can write $\psi_2$ and $\psi_3$ in term of $\psi_1$ as follows
\begin{eqnarray}\label{si2}\label{si3}
 \psi_2&=&\frac{1}{2}\left(1+3\not x\right)\left[\not
Z+(1+3\not x)x\cdot Z\right]\psi_1, \\\label{si3}
 \psi_3&=&-\frac{5}{4}\left(1-\not x\right)\left[\not
Z+(1+3\not x)x\cdot Z\right]\psi_1.
\end{eqnarray}
Using the divergencelessness condition and the above relations, one obtains
\begin{eqnarray}  \label{confpola}
 \left(Z\cdot \partial^T+3x\cdot Z\right) \psi_1+\frac{1}{2}(\not x-3)\left[ \not Z +(1+3\not x)x\cdot Z\right] \psi_1=0,\end{eqnarray}
and making use of $\not \Psi=0$, leads one to write
 $$ \not Z^T\psi_1-\left(4\not x +\not \partial^T\right)\psi_2+4\psi_3=0,$$
and in terms of $\psi_1$, this equation becomes
 \b \label{gama=0}
 \not Z^T\psi_1+2(\not x +5)\left[ \not Z+(1+3\not x)x\cdot Z\right] \psi_1-\frac{2}{5}(1+2\not x) x\cdot Z\psi_1=0 .
  \e
Actually what we have obtained is that by dealing with $\psi_1$, the other two fields will
be established as well. Consequently by gathering all the results, $\Psi$ can be written as follows \b
\Psi_\alpha (x)=\textbf{D}_\alpha(x,\partial^T,Z)\psi_1,\e
where we have defined
$$\textbf{D}_\alpha(x,\partial^T,Z)\equiv Z^T_\alpha+\left[\frac{1}{2}{\cal D}^{(\frac{3}{2})}_\alpha\left(1+3\not x\right)-\frac{5}{4}\gamma^T_\alpha(1-\not x)\right]\left[\not
Z+(1+3\not x)x\cdot Z\right].$$
In this formalism a given spin-$\frac{3}{2}$ field could be constructed from the multiplication of the polarization vector $\textbf{D}_\alpha$ or $Z_\alpha$ ($5$ degrees of freedom) with the spinor field $\psi_1$ ($2$ degrees of freedom), which appears naturally $10$ polarization states. After making use of (\ref{confpola}) and (\ref{gama=0}), the degrees of freedom are indeed reduced to the usual $4$ polarization states $m_{\frac{3}{2}}=\frac{3}{2},\frac{1}{2},-\frac{1}{2}$ and $-\frac{3}{2}$, where two of them are the physical states $\pm\frac{3}{2}$ \cite{ta1403}.

Now, $\psi_1$ should be identified. Making use of the relation between the $s=1/2$ and spin-zero Casimir operators, equation (\ref{psi1}) can be written as
\begin{equation} \left( Q_{\frac{1}{2}}^{(1)}-\frac{1}{2}\right)\psi_1(x)=0. \end{equation}
This means that $\psi_1$ and its related two-point functions can in fact be extracted from a massive spinor field in the principal series representation given by (\ref{equation}) by setting $\nu=-i$. Massive spinor field and its two-point functions has already been studied in \cite{sep}. Therefore, the solution of (\ref{psi1}) are found to be
 \begin{eqnarray}
\psi_{1}(x)&=&(x\cdot\xi
)^{-2}{\cal V}(x,\xi), \nonumber\\
 \psi^{'}_{1}(x)&=&(x\cdot\xi)^{-1}{\cal U}(\xi),
\end{eqnarray}
${\cal V}(x,\xi)\equiv\frac{\not
x \not \xi}{x\cdot\xi}{\cal V}(\xi)$, these solutions are the de Sitter-Dirac plane waves. $\xi$ is a vector that
lives in the positive sheet of the light cone, \textit{i.e.,}
$$\xi \in \C^+=\{ \xi \;\;;\eta_{\alpha \beta}\xi^\alpha
\xi^\beta=(\xi^0)^2-\vec \xi\cdot\vec \xi-(\xi^4)^2=0,\; \xi^0>0 \}.$$
$\xi$ plays the role of the energy momentum in the null curvature limit. Two spinors ${\cal V}(\xi)$ and ${\cal U}(\xi)$ are given by \cite{sep}
\b {\cal U}^i(\xi)=\frac{\xi^0-\vec \xi\cdot \vec \gamma
 \gamma^0+1}{\sqrt{2(\xi^0+1)}}{\cal U}^i(\stackrel{o}{\xi}_+),\;\;\; {\cal V}^i(\xi)=\frac{1}{\sqrt{2(\xi^0+1)}}{\cal U}^i(\stackrel{o}{\xi}_-), \;\;i=1,2, \e
 where
  \begin{eqnarray}
  {\cal U}_1(\stackrel{o}{\xi}_+) &=&\frac{1}{\sqrt{2}}\left( \begin{array}{clcr} \chi\\ \chi \\ \end{array} \right) ,\hspace{6mm} {\cal U}_2  (\stackrel{o}{\xi}_+)=\frac{1}{\sqrt{2}}\left( \begin{array}{clcr} \chi'\\ \chi' \\ \end{array} \right),\nonumber\\
 {\cal U}_1(\stackrel{o}{\xi}_-) &=&\frac{1}{\sqrt{2}}\left(
\begin{array}{clcr} \chi\\-\chi \\ \end{array} \right) ,\hspace{6mm}
{\cal U}_2 (\stackrel{o}{\xi}_-)=\frac{1}{\sqrt{2}}\left(
\begin{array}{clcr} \chi'\\-\chi' \\ \end{array} \right),
\end{eqnarray}
in which $\chi=\left(
\begin{array}{clcr}1 \\ 0 \\ \end{array} \right)$ , $\chi'=\left(
\begin{array}{clcr} 0 \\ 1 \\ \end{array} \right)$ and $ \xi
=\stackrel{o}{\xi}_\pm \equiv (1,\vec 0,\pm 1)$.

Therefore we have two solutions for $ \Psi_\alpha(x)$ which are as follows
\b
 \Psi_{1\alpha}(x)=\textbf{D}_\alpha(x,\partial^T,Z){\cal V}(x,\xi) (x\cdot\xi )^{-2}\equiv {\cal
V_\alpha}(x,\xi,Z)(x\cdot \xi )^{-2},
\e
and
\b
\Psi_{2\alpha}(x)=\textbf{D}_\alpha(x,\partial^T,Z){\cal U}(\xi)(x\cdot\xi)^{-1}\equiv  {\cal U_\alpha}(x,\xi,Z)(x\cdot \xi)^{-1},
\e
After doing some calculation and taking the derivative, the explicit form of ${\cal U}_\alpha$ and
${\cal V}_\alpha$ are obtained in terms of $x, \xi$ and $Z$.

\setcounter{equation}{0}
\section{Two-point function in ambient space formalism}

In this section, we deal with conformally invariant two-point functions of a massless
spin-$\frac{3}{2}$ field. They are found in terms of the spinor (spin-1/2) two-point function. The two-point
function of massless spin-$\frac{3}{2}$ field is given by
\b\label{two} S^{i \bar
j}_{\alpha\alpha'}(x,x')=<\Omega\mid\Psi^i_\alpha(x){\overline
\Psi}^{\overline j}_{\alpha'}(x')\mid\Omega>,\e where
$x,x' \in X_H$, and $|\Omega> $ stands for the vacuum state.
Let us use the similar procedure of the previous section and write
the desired two-point function in terms of three spinor two-point
functions and impose some conditions to write down the two of them
as a function of the third one. By using the equations (\ref{6.1}) and (\ref{two}), the following form of two-point function is proposed \b {\cal S}_{\alpha
\alpha'}(x,x')=\theta_{\alpha}\cdot \theta'_{\alpha' } {\cal
S}_1(x,x')-{\cal D}^{(\frac{3}{2})}_{\alpha} {\cal
S}_2(x,x'){\gamma^4}\overleftarrow{{\cal
D}}'^{(\frac{3}{2})}_{\alpha'}{\gamma^4} -\gamma^T_\alpha {\cal
S}_3(x,x'){\gamma^4}{\gamma'}^T_{\alpha'}{\gamma^4},\e
where the prime operators act only on $x'$.  This two-point
function must satisfy the CI system of the field equations (\ref{4.212}). If one demands that ${\cal S}_{\alpha
\alpha'}(x,x')$ satisfies the second order field equation and using the identities of appendix B, the following relations are obtained
 \begin{eqnarray}
 &(Q_0+\not x\not\partial^T-3){\cal S}_1(x,x')=0,\label{7.3}\\
 &2(x\cdot \theta'_{\alpha' }){\cal S}_1
(x,x')+(Q_0+\not
x\not\partial^T){\cal S}_2(x,x'){\gamma^4}\overleftarrow{{\cal D}}'^{(\frac{3}{2})}_{\alpha'}{\gamma^4}=0,\label{7.4}\\\label{7.5}
&\left[-3\not x (x\cdot\theta'_{\alpha' })+(\theta'_{\alpha'
}\cdot\gamma^T_{\alpha})\right]{\cal S}_1(x,x')-(Q_0+\not
x\not\partial^T){\cal
S}_3(x,x'){\gamma^4}\gamma'^T_{\alpha'}{\gamma^4}=0.
\end{eqnarray}
On the other hand, ${\cal S}_{\alpha\alpha'}(x,x')$ must satisfy the first order equation, then one obtains
\begin{equation}
\begin{array}{l}  \label{7.6}
(\not x\not\partial^T-1){\cal S}_1(x,x')=0,\\
(\not x -2){\cal
S}_2(x,x'){\gamma^4}\overleftarrow{{\cal
D}}'^{\frac{3}{2}}_{\alpha'}{\gamma^4}-2\not x{\cal
S}_3(x,x'){\gamma^4}\gamma'^T_{\alpha'}{\gamma^4}=0,\\
\not x \not\partial^T{\cal
S}_3(x,x'){\gamma^4}\gamma'^T_{\alpha'}{\gamma^4}-(\not
x +4)
{\cal S}_2(x,x'){\gamma^4}\overleftarrow{{\cal D}}'^{\frac{3}{2}}_{\alpha'}{\gamma^4}+x_{\alpha'} \not x {\cal S}_1(x,x')=0.
\end{array}
\end{equation}
We can write $ {\cal S}_2(x,x') $ in term of $ {\cal S}_3(x,x')$ and vice-versa by using the second equation (\ref{7.6}) as fallows
$$
{\cal S}_2(x,x'){\gamma^4}\overleftarrow{{\cal D}}'^{\frac{3}{2}}_{\alpha'}{\gamma^4}=\frac{2}{5}(1-2\not x ){\cal
S}_3(x,x'){\gamma^4}\gamma'^T_{\alpha'}{\gamma^4},$$
\begin{equation} \label{ss}{\cal S}_3(x,x'){\gamma^4}\gamma'^T_{\alpha'}{\gamma^4}=\frac{1}{2}(1+2\not x){\cal S}_2(x,x'){\gamma^4}\overleftarrow{{\cal D}}'^{\frac{3}{2}}_{\alpha'}{\gamma^4}.
\end{equation}
The first order equation of two-point function can be written as
$$(\not x\not\partial^T-1){\cal S}_1(x,x')=0,$$
$$(\not x +2)\not \partial^T {\cal
S}_2(x,x'){\gamma^4}\overleftarrow{{\cal
D}}'^{\frac{3}{2}}_{\alpha'}{\gamma^4}+2(\not x-8) {\cal
S}_2(x,x'){\gamma^4}\overleftarrow{{\cal
D}}'^{\frac{3}{2}}_{\alpha'}{\gamma^4}+2x_{\alpha'}\not x {\cal S}_1(x,x') =0,$$\b \label{7.8} \not x \not\partial^T{\cal
S}_3(x,x'){\gamma^4}\gamma'^T_{\alpha'}{\gamma^4}-\frac{2}{5}(6-7\not x){\cal
S}_3(x,x'){\gamma^4}\gamma'^T_{\alpha'}{\gamma^4}+x_{\alpha'}\not x {\cal S}_1(x,x')=0.  \e
The equation governing ${\cal S}_1$ can be deduced from the equation (\ref{7.3}) and the first equation of (\ref{7.6}) as
\b
\left(Q_0-2\right)S_1(x,x')=0,\;\;\left(\not x\not\partial^T-1\right){\cal S}_1(x,x')=0. \e
In Ref. \cite{sep},  the solutions have been obtained, here, we only quote the result \b \label{11.12}
S_1(x,x')=\frac{i}{2(4\pi)^2}\left[ P^{(7)}_{-3}(x\cdot x')\not x
-3P^{(7)}_{-1}(x\cdot x')\not x'\right] \gamma^{4},\e $P^{(7)}_\sigma$
are the generalized Legendre functions (see appendix D). Similar
to the previous section, by acting $\not \partial^T$ on the first order equation (\ref{7.8}), one finds the second order equation for ${\cal S}_2$ and ${\cal S}_3$ as follows
\b \label{new3}
Q_0 {\cal
S}_2(x,x'){\gamma^4}\overleftarrow{{\cal
D}}'^{\frac{3}{2}}_{\alpha'}{\gamma^4}=-2(3\not x+20)  {\cal
S}_2(x,x'){\gamma^4}\overleftarrow{{\cal
D}}'^{\frac{3}{2}}_{\alpha'}{\gamma^4}+\frac{2}{5}\left[2x_{\alpha'}(1+7\not x )+(1-2\not x)\gamma^T_{\alpha'}\right]{\cal S}_1(x,x'), \e
\b \label{new4}
Q_0{\cal
S}_3(x,x'){\gamma^4}\gamma'^T_{\alpha'}{\gamma^4}=\frac{2}{5}(7\not x-16) {\cal
S}_3(x,x'){\gamma^4}\gamma'^T_{\alpha'}{\gamma^4}+\left[\frac{2}{5} x_{\alpha'}(6\not x -7)+\gamma^T_{\alpha'}\right]{\cal S}_1(x,x').\e
After making use of the above relations in (\ref{7.5}), one can write ${\cal S}_2$ and ${\cal S}_3$ in
terms of ${\cal S}_1$ as follows \b {\cal
S}_2(x,x'){\gamma^4}\overleftarrow{{\cal D}}'^{\frac{3}{2}}_{\alpha'}{\gamma^4}= -\frac{1}{10}(1-2\not x)\left[-3 \not x (x.\theta'_{\alpha' })+(\theta'_{\alpha'}.\gamma^T_{\alpha})-\frac{1}{5} x_{\alpha'}(7\not x-14)-\gamma^T_{\alpha'}\right]{\cal S}_1(x,x'),\e
\b {\cal S}_3(x,x'){\gamma^4}\gamma'^T_{\alpha'}{\gamma^4}= -\frac{1}{4}\left[-3 \not x (x.\theta'_{\alpha' })+(\theta'_{\alpha'}.\gamma^T_{\alpha})- \frac{1}{5}x_{\alpha'}(7\not x-14)-\gamma^T_{\alpha'}\right]{\cal S}_1(x,x'). \e
Finally the two-point function takes the following form in terms of ${\cal
S}_1$
 \b \label{11.11} {\cal S}_{\alpha
\alpha'}(x,x')= \textbf{D}_{\alpha \alpha'}(x, \partial^T ;x',
{\partial'}^T) {\cal S}_1(x,x'),\e where we have
 $$
\textbf{D}_{\alpha \alpha'}(x,
\partial^T ;x', {\partial'}^T)=\theta_{\alpha}\cdot \theta'_{\alpha' }+\left(\frac{1}{10}{\cal D}^{(\frac{3}{2})}_{\alpha}(1-2\not x)+\frac{1}{4}\gamma^T_\alpha\right)$$ \b \times \left[-3 \not x (x.\theta'_{\alpha' })+(\theta'_{\alpha'}.\gamma^T_{\alpha})-\frac{1}{5} x_{\alpha'}(7\not x-14)-\gamma^T_{\alpha'}\right].\e This
two-point function is conformally invariant.

\setcounter{equation}{0}
\section{Two-point functions in intrinsic coordinates}

The two-point function in intrinsic coordinates of de Sitter space can indeed be obtained by projecting ${\cal S}_{\alpha,\alpha'}$. To do this let us recall briefly the de Sitter-Dirac equation in embedding and intrinsic spaces. In embedding space the $ y^\alpha\equiv( y^\mu,y^4=H^{-1})$ is introduced with the following relation with $x^\alpha$ \cite{Gur}

$$ x^\alpha=\left(Hy^4\right)f^\alpha\left( y^0,y^1,y^2,y^3\right),$$
where arbitrary function $ f^\alpha$ satisfying $f^\alpha
f_\alpha=-H^{-2}$. The five matrices $\beta^\alpha\equiv(\dfrac{\partial y^\alpha}{\partial x^\beta})\gamma ^\beta $ then satisfy the anticommutation relations
$$ \left\{ \beta^\mu ,\beta^\nu \right\}=2g^{\mu\nu}, \hspace*{3mm} \left\{\beta^\mu, \beta^4\right\}=0,$$
with $ g^{\mu\nu}=\dfrac{\partial y^\mu}{\partial x^\alpha}\dfrac{\partial y^\nu}{\partial x^\beta}\eta^{\alpha\beta}$. Suppose that $\psi(x)$ be a solution of (\ref{dirac equation}), if one writes $\psi=(1\pm i\beta^4)\chi$, then $\chi$ satisfies the following equation
\b\label{11.1} \Big( \beta^\mu \dfrac{\partial}{\partial y^\mu}-2H\beta^4-m\Big)\chi=0, \e
this is the G\"ursey-Lee equation \cite{Gur}. This equation turns to the usual Dirac equation in de Sitter space by choosing a local vierbein $ e^\alpha_{\mu}$ at every point of de Sitter space and by setting $ \gamma^\mu (X)=V\beta^\mu(y)V^{-1}$, then under the $V$ transformation, one obtains
$$\Big( \gamma^{\mu}(X){\nabla}_{\mu}-m\Big)\Psi(X)=0. $$
Note that $\beta^4=\gamma_{\alpha}x^{\alpha}$ is related to the constant matrix $\gamma^4$ by $\gamma^4=V\beta^4V^{-1}$.

Any linear combination of $(1+i\beta^4)\chi$ and $(1-i\beta^4)\chi$ also satisfies the G\"ursey-Lee equation
\b\label{11.2} \psi(x)=\dfrac{a_1}{2}(1+i\beta^4)\chi + \dfrac{a_2}{2}(1-i\beta^4)\chi,\e where the coefficients $a_1$ and $a_2$ are fixed by the normalization condition.

Now the relation between the spinor fields defined in embedding space and intrinsic space can be written as
\b\label{11.3} \psi(x)=\dfrac{a_1}{2}V^{-1}(1+i\gamma^4)\Psi(X) + \dfrac{a_2}{2}V^{-1}(1-i\gamma^4)\Psi(X),\e
or equivalently,
\b\label{11.4}\Psi(X)=\left[\dfrac{1}{2a_1}(1+i\gamma^4)+\dfrac{1}{2a_2}(1-i\gamma^4)\right]V\psi(x),\e
Similarly, for spin-$\dfrac{3}{2}$ field one obtains
\begin{eqnarray}\label{{11.5}}
 \Psi_{\mu}(X)&=&\left[ \dfrac{1}{2a_1}(1+i\gamma^4)+\dfrac{1}{2a_2}(1-i\gamma^4)\right]V\dfrac{\partial x^{\alpha}}{\partial X^{\mu}}\psi_{\alpha}(x),\\ \nonumber
&\equiv & {\cal A}\dfrac{\partial x^{\alpha}}{\partial X^{\mu}}\psi_{\alpha}(x),
\end{eqnarray}
where ${\cal A}$ matrix is given by
$$ {\cal A}=\left[ \dfrac{1}{2a_1}(1+i\gamma^4)+\dfrac{1}{2a_2}(1-i\gamma^4)\right]V,$$
 In this case, the covariant derivatives in de Sitter intrinsic space are related to the derivative in embedding space as \cite{me}
\begin{equation}
\begin{array}{c}
\label{11.6} \nabla_{\mu}\Psi_{\nu}(X)={\cal A}\dfrac{\partial x^{\alpha}}{\partial X^{\mu}}\dfrac{\partial x^{\beta}}{\partial X^{\nu}}\left(\partial^T_{\alpha}\psi_{\beta}- x_{\beta}\psi_{\alpha}\right),\\
\nabla_{\mu}\nabla_{\nu}\Psi_{\rho}(X)={\cal A}\dfrac{\partial x^{\alpha}}{\partial X^{\mu}}\dfrac{\partial x^{\beta}}{\partial X^{\nu}}\dfrac{\partial x^{\gamma}}{\partial X^{\rho}}Trpr \partial^T_{\alpha}Trpr\partial^T_{\beta}\psi_{\gamma} \\
={\cal A}\dfrac{\partial x^{\alpha}}{\partial X^{\mu}}\dfrac{\partial x^{\beta}}{\partial X^{\nu}}\dfrac{\partial x^{\gamma}}{\partial X^{\rho}}\left[\partial^T_{\alpha}(\partial^T_{\beta}\psi_{\gamma}-x_{\gamma}\psi_{\beta})-x_{\beta}(\partial^T_{\alpha}\psi_{\gamma}-x_{\gamma}\psi_{\alpha})-x_{\gamma}(\partial^T_{\beta}\psi_{\alpha}-x_{\alpha}\psi_{\beta})\right],
\end{array}
\end{equation}
$Trpr$ is the abbreviation of the transverse projection. Therefore, the two-point function in de Sitter space is related to the embedding space one as
\b\label{11.8}W_{\mu\mu'}=\dfrac{\partial x^{\alpha}}{\partial X^{\mu}}\dfrac{\partial x'^{\alpha'}}{\partial X'^{\mu'}}{\cal A}{\cal S}_{\alpha\alpha'}{\cal A}^t,\e
in this equation ${\cal S}_{\alpha\alpha'}$ is defined by equation (\ref{11.11}) and ${\cal A}^t$ is
$$ {\cal A}^t={\cal A}^\dagger\gamma^{0}\gamma^{4}=\left[\dfrac{1}{2a_1}(1+i\gamma^4)+\dfrac{1}{2a_2}(1-i\gamma^4)\right]V^t.$$

On the other hand, one can use three basic tensors to expand any maximally symmetric bitensor, these tensors are defined below \cite{ta1}
$$n_{\mu}=\nabla_{\mu}z(x,x')\;\;\;,\;\;\;n_{\mu'}=\nabla_{\mu'}z(x,x'),$$
and the parallel propagator
$$ g_{\mu\nu'}=-c^{-1}({\cal P})\nabla_{\mu}n_{\nu'}+n_{\mu}n_{\nu'}.$$
The coefficients in this expansion are functions of the geodesic distance $z(x,x')$, between the points $x$ and $x'$.  It is de Sitter invariant and can be defined by an unique analytic extension. In this senses, these fundamental tensors form a complete set.

If ${\cal P}\equiv-x\cdot x'$, then one can write
 \b\left\{\begin{array}{c}
  {\cal P}=\cosh z,\hspace{2mm}\mbox{if $x$ and $x'$ are timelike separated,}\\
  {\cal P}=\cos z,\hspace{4mm}\mbox{if $x$ and $x'$ are spacelike separated.}\\
\end{array}
 \right.\e
In ambient space notations these tensors become
$$ \partial^T_{\alpha}z(x, x'),\;\;\;\partial'^T_{\beta'}z(x, x'),\;\;\;\theta_{\alpha}.\theta'_{\beta'}.$$
For ${\cal P}=\cos z$, one can find
$$n_{\mu}=\dfrac{\partial x^{\alpha}}{\partial X^{\mu}}\partial^T_{\alpha}z(x,x')=\dfrac{\partial x^{\alpha}}{\partial X^{\mu}}\dfrac{(x'\cdot \theta_{\alpha})}{\sqrt{1-{\cal P}^2}},$$
$$n_{\nu'}=\dfrac{\partial x'^{\beta'}}{\partial X'^{\nu'}}\partial^T_{\beta'}z(x,x')=\dfrac{\partial x'^{\beta'}}{\partial X'^{\nu'}}\dfrac{(x\cdot \theta'_{\beta'})}{\sqrt{1-{\cal P}^2}},$$
$$\nabla_{\mu}n_{\nu'}=\dfrac{\partial x^{\alpha}}{\partial X^{\mu}}\dfrac{\partial x'^{\beta'}}{\partial X'^{\nu'}}{\theta}_{\alpha}^{\varrho} {\theta'}_{\beta'}^{\gamma'}\partial^T_{\varrho}\partial^T_{\gamma'}z(x,x')=c({\cal P})\left[n_{\mu}n_{\nu'}{\cal P}-\dfrac{\partial x^{\alpha}}{\partial X^{\mu}}\dfrac{\partial x'^{\beta'}}{\partial X'^{\nu'}}{\theta}_{\alpha}\cdot {\theta'}_{\beta'}\right],$$
where $c^{-1}({\cal P})=-\dfrac{1}{\sqrt{1-{\cal P}^2}}$. In the case of ${\cal P}=\cosh z,$ one has $c({\cal P})=-\dfrac{i}{\sqrt{1-{\cal P}^2}}$, and hence in both cases, one obtains
$$g_{\mu\nu'}+({\cal P}-1)n_{\mu}n_{\nu'}=\dfrac{\partial x^{\alpha}}{\partial X^{\mu}}\dfrac{\partial x'^{\beta'}}{\partial X'^{\nu'}}{\theta}_{\alpha}\cdot {\theta'}_{\beta'}.$$
After making use of the above relations and (\ref{11.8}), the two-point function (\ref{11.11}) in de Sitter space is found as
$$ \label{11.9} W_{{\mu}{\mu'}}={\cal A}\left\{g_{{\mu}{\mu'}}+({\cal P}-1)n_{\mu}n_{\mu'}+\left(\frac{1}{10}(-\nabla_{\mu}-\gamma^T_{\mu}(X)\not x)(1-2\not x)+\frac{1}{4} \gamma^T_{\mu}(X)\right)\right.$$
\b \left. \times\left[-3\not x(X\cdot g'_{\mu'})+( g'_{\mu'}.\gamma^T_{\mu}(X))-\frac{1}{5} X_{\mu'}(7\not x-14)-\gamma^T_{\mu'}(X)\right]\right\}{\cal S}_1 {\cal A}^{t},\e where ${\cal S}_1$ is given by (\ref{11.12}). This two-point function is obviously dS-invariant.

\section{Conclusions}

Conformal theories and conformal techniques have been used for a long time in physics (see for example \cite{Nakayama} and references therein), e.g., in the gravitational physics, the main motivation of considering such theories is to achieve a proper theory of quantum gravity. As one knows, the quantum theory of gravity based on Einstein equation is not renormalizable \cite{brill}, however, it is proved that the conformal theories of gravity are better to renormalize \cite{Stelle}. On the other hand, the gravitational field is long range and propagates with the speed of light, thus in the linear approximation, it is expected that the equations governing its dynamic must be conformally invariant, whereas, the Einstein equation is not conformally invariant equation. Moreover, within the standard theory of gravity one needs to introduce a huge amount of dark and non-detectable matter and energy to explain some extraordinary phenomena \cite{dark1}, while, the conformally invariant Weyl gravity can address some of these large scale issues.

On the other hand as it is known, the powerful $AdS/CFT$ conjecture makes a relation between the quantum aspects of gravity in anti-de Sitter space and dual conformal theory which lives on its boundary. But as mentioned our universe governed by a non-vanishing positive cosmological constant and hence, a proper dictionary of $dS/CFT$ is desired. It is believed that a better understanding of higher-spin and gauge theories in de Sitter space will help one to find a proper $dS/CFT$ correspondence \cite{stro}.

The massless spin-$\frac{3}{2}$ (gravitino) field is supposed to be the fermionic partner of gravitational field which is used in supergravity to unify gravitational (spin-2) and nongravitational (spin-1) forces \cite{ta5}. In this paper by choosing the de Sitter space as the background, and with the aid of the background symmetries which well appear in dS group, we considered the massless spin-$\frac{3}{2}$ field by constructing its conformal invariant field equation. The solutions and two-point functions are also found.

\vspace{0.5cm}
\section*{Acknowledgements}
We would like to thank referees for their comments and criticisms that resulted in improvement of the manuscript. We are grateful to  M.R. Enaiati, M. Parsamehr for useful discussions. MRT would also like to acknowledge M. Tanhayi for comments on higher-spin fields.
\appendix

\section{a note on UIR of de Sitter group}

The unitary irreducible representation of the dS group are classified
according to a pair $(p,q)$ of the possible spectral values of the
Casimir operators as follows \cite{dixmier}
\begin{equation}
<Q^{(1)}_{p}>=\left(-p(p+1)-(q+1)(q-2)\right)\1 ,\qquad\quad
<Q^{(2)}_{p}>=\left(-p(p+1)q(q-1)\right)\1.
\end{equation}
For each kind of field, the possible
range of the parameters $q$ and $p$ (where $2p\in  \N$ and
$q\in \C$) indicates the unitary and also irreducible
representation, generally they characterize three types of UIR in
dS, which can be listed as follows\\
\begin{enumerate}
    \item  Principal series $U_{p,q}$ \begin{equation}
\left\{\begin{array}{ll} p=s,\hspace{2mm}\,\,\,\,\mbox{and} \,\,\,q=\frac{1}{2}+i\nu,\,\,\,\,\mbox{for}\,\,\,\,\,\,\,\nu\geq 0,\,\,\,\,\,s=0,1,2,\cdots,\\
p=s,\,\,\,\,\,\,\,\mbox{and}
\,\,\,\,q=\frac{1}{2}+i\nu,\,\,\,\,\,\mbox{for}\,\,\,\,\,\,\,\nu>
0,\,\,\,\,\,s=1/2,3/2\cdots.
\end{array}\right.
\end{equation}
which also called the massive representation because in the flat
limit they tend to the massive representation of Poincar\'e group
with spin $s$.
  \item Complementary series $V_{p,q}$
 \begin{equation}  \left\{\begin{array}{ll} p=s,\hspace{2mm}\,\,\,\,\mbox{and} \,\,\,q=\frac{1}{2}+\nu,\,\,\,\,\mbox{for}\,\,\,\,\,\,\,0 < |\nu|<3/2 ,\,\,\,\,\,s=0,\\
p=s,\,\,\,\,\,\,\,\mbox{and}
\,\,\,\,q=\frac{1}{2}+\nu,\,\,\,\,\,\mbox{for}\,\,\,\,\,\,\,0<|\nu|<1/2,\,\,\,\,\,s=1,2,\cdots.
\end{array}\right.
\end{equation}
in this case $\nu\in \R$, and the massless conformally
coupled scalar field in this series has the Poincar\'e limit. \item
Discrete series $\Pi^\pm_{p,q}$: in this case the only representation
which has the Poincar\'e limit is $p=q=s$. This series represents the
massless fields.
\end{enumerate}
In dS space the massless field with spin-$\frac{1}{2}$ associates with
discrete series $\prod^{\pm}_{p,s}$ and their UIRs are
$\prod^{\pm}_{\frac{1}{2},\frac{1}{2}}$, where $p=q=s=\frac{1}{2}$
correspond to:$<Q^{(1)}_{\frac{1}{2}}>=\frac{3}{2}$ and these two
representations have a Minkowskian interpretation. For
spin-$\frac{3}{2}$ field, the two UIRs
$\Pi^{\pm}_{\frac{3}{2},\frac{3}{2}}$, have a Minkowskian
interpretation. For example $\Pi^{\pm}_{\frac{3}{2},\frac{1}{2}}$
have no corresponding flat limit. It is proved that for every
massless representation of Poincar\'e group there exists only one
corresponding representation in the conformal group \cite{san,
meh}. In the massless case, 
conformal invariance leads one to deal with the discrete
series representations and their lower limits of the universal
covering of the conformal group. The conformal group,
is locally isomorphic to
$SO(2,4)$ and its compact subgroup is $SO(2)\times SO(4)$, where 
the generator of $SO(2)$ is the conformal energy. The unitary irreducible representation of the conformal group are
denoted in the sequel by ${\cal C}(\pm E_0,j_1,
j_2)$, where $(j_1,j_2) \in\N/2 \times \N/2$ labels the UIRs of
$SU(2) \times SU(2)$ and $E_0$ stands for the positive (resp.
negative) conformal energy. The direct sum of two UIRs $C(j+1,j,0)$ and $C(-j-1,j,0)$
of the conformal group with positive and negative energy and
$j=\frac{1}{2}$ for spinor field $j=\frac{3}{2}$ for
spin-$\frac{3}{2}$ field, is a unique extension of representation
$\prod^{+}_{j,j}$. We shall denote the massless Poincar\'e UIRs by $P^{>}(0,j)$ and
$P^{<}(0,j)$ which respectively are the positive and negative energies representation with 
positive helicity. The following diagrams illustrate these
relations
 \b
\left.
\begin{array}{ccccccc} &             & {\cal C}(j+1, j,0)
& &{\cal C}(j+1, j,0)   &\hookleftarrow &{\cal P}^{>}(0,j)\\
\Pi^+_{j,j} &\hookrightarrow  & \oplus
&\stackrel{H=0}{\longrightarrow} & \oplus  & &\oplus  \\
 &             & {\cal C}(-j-1, j,0)&
& {\cal C}(-j-1, j,0)  &\hookleftarrow &{\cal P}^{<}(0,j),\nonumber\\
\end{array} \right. \e
\b \left. \begin{array}{ccccccc}
 &             & {\cal C}(j+1, 0,j)
& &{\cal C}(j+1, 0,j) &\hookleftarrow &{\cal P}^{>}(0,-j)\\
\Pi^-_{j,j} &\hookrightarrow  & \oplus
&\stackrel{H=0}{\longrightarrow} &  \oplus & &\oplus  \\
&             & {\cal C}(-j-1, 0,j)&
& {\cal C}(-j-1, 0,j)   &\hookleftarrow &{\cal       P}^{<}(0,-j),\\
\end{array}\nonumber \right. \e
where the arrows $\hookrightarrow $ designate unique extension.

\section{ Some useful relations}

In this appendix, some useful relations are given which are used
in this paper

\begin{equation}
\begin{array}{llcr}\nonumber
[\partial^T_\alpha , \partial^T_\beta]=x_\beta\partial^T_\alpha-x_\alpha\partial^T_\beta,&
[\partial^T_\alpha , x_\beta]=\theta_{\alpha\beta},\\

[\gamma^T_{\alpha}, x_\beta]=0,&
[\gamma^T_{\alpha}, \partial^T_\beta]=-(\theta_{\alpha\beta}\not x + x_\alpha \gamma^T_{\beta}),\\
 Q_{0}=-{(\partial^T_\alpha)}^2,&
Q_{0}{\partial^T_{\alpha}}={\partial^T_{\alpha}}Q_{0}+2{\partial^T_{\alpha}}+2x_{\alpha}Q_{0},\\
Q_{0}x_{\alpha}=x_{\alpha}Q_{0}-4x_{\alpha}-2{\partial^T_{\alpha}},&
{\gamma^T_{\alpha}}=\Theta_{\alpha}^{\beta}\gamma_{\beta}=\gamma_{\alpha}+x_{\alpha}x\cdot \gamma,\\
Q^{(1)}_{\frac{3}{2}}{\cal D}^{(\frac{3}{2})}_\alpha={\cal D}^{(\frac{3}{2})}_\alpha Q^{(1)}_{\frac{1}{2}},&
 \gamma^T_\alpha\not x=-\not
x \gamma^T_\alpha,\\
Q_\frac{3}{2}^{(1)}\partial^T_\alpha=\partial^T_\alpha(Q_0+\not
x\not\partial^T),&
 \not\partial^T
{\cal D}^{(\frac{3}{2})}_{\alpha}={\cal
D}^{(\frac{3}{2})}_{\alpha}\not\partial^T-\not
x\partial^T_\alpha-x_\alpha\not\partial^T-4x_\alpha\not x +
3\gamma^T_\alpha,\\
\not x\not\partial^T x_\alpha=x_\alpha\not
x\not\partial^T+\not x\gamma^T_\alpha,&
 Q_0\not x\not\partial^T=\not
x\not\partial^T Q_0,\\
 Z^T_\alpha=Z_\alpha+x_\alpha x\cdot Z, &
  Q_0 Z^T_\alpha=Z^T_\alpha Q_0-2x_\alpha
Z\cdot \partial^T-4x_\alpha x\cdot Z,\\
\partial^T\cdot Z^T=Z\cdot \partial^T+4x\cdot Z,&
 \not Z^T=\not Z+\not x x\cdot Z,\\
\not\partial^T(Z\cdot \partial^T)=-2\not x (Z\cdot \partial^T)+\not
x(x\cdot Z),& \not
\partial^T(x\cdot Z)=-\not x(x\cdot Z)+\not Z^T,\\ (x\cdot \theta'_{\alpha'})(x\cdot \theta'^{\alpha'})=(x\cdot x')^2-1,&\partial^T_{\alpha}(x\cdot \theta'_{\beta'})=\theta_{\alpha}\cdot \theta'_{\beta'},\\ \partial^T_{\alpha}(x'\cdot \theta_{\beta})=x_{\beta}(x'\cdot \theta_{\alpha})-Z\theta_{\alpha\beta},& \partial^T_{\alpha}(\theta_{\beta}\cdot \theta'_{\beta'})=x_{\beta}(\theta_{\alpha}\cdot \theta'_{\beta'})+\theta_{\alpha\beta}(x\cdot \theta'_{\beta'}),\\
x_\alpha(Z.\partial^T)=(Z.\partial^T)x_\alpha-Z^T_\alpha,

\end{array}
\end{equation}
and also the following identities are used
\begin{equation}
\begin{array}{l}\nonumber
Q_0\not
x(Z\cdot \partial^T)=\not x(Z\cdot \partial^T)Q_0-2\not
x(Z\cdot \partial^T)+2\not x (x\cdot Z)Q_0-2\not
\partial^T(Z\cdot \partial^T),\\
 \not\partial^T\not Z^T=\not Z^T\not\partial^T+\not
x\not Z^T+4x\cdot Z,
 Q_0\not x(x\cdot Z)=\not x x\cdot Z Q_0-8\not x
x\cdot Z-2\not x(Z\cdot \partial^T)-2\not\partial^T(x\cdot Z),\\
Q_0{\cal D}^{(\frac{3}{2})}_{\alpha}={\cal
D}^{(\frac{3}{2})}_{\alpha}Q_0-4\partial^T_\alpha+4\gamma^T_\alpha\not
x-2\gamma^T_\alpha\not\partial^T-2x_\alpha\not
x\not\partial^T\cdot
\end{array}
\end{equation}

\section{Conformally invariant wave equation}

In this appendix we give some details of CI equations. As it was
shown, we obtained the following equation from (\ref{4.2}) by choosing $n=1$
\b\label{eq}(Q_{0}-2)\Phi_{\alpha}=0,\e note $\Phi_\alpha$ is
defined on conformal space. Multiply (\ref{eq}) by $x_{\alpha}$ from
the left and after making use of the relations given in Appendix
B, leads one to write
\b Q_{0}x\cdot \Phi+2x\cdot\Phi+2\partial^T\cdot\Phi=0.\e
 The transversality condition on cone yields
$$ u_a\Phi^{ab...}=0=u_5\Phi^5+u_\alpha\Phi^\alpha,$$
$$ x_5(\Phi^5+x.\Phi)=0,$$ then by acitng $(Q_0-2)$ on this relation, we obtain
\b (Q_{0}-2)x\cdot \Phi=0. \e
By using (C.3) in (C.2) one has \b \label{c4} {\partial}^T\cdot \Phi=-2x\cdot \Phi.\e The divergence of
$\Psi_{\alpha}$ defined in (\ref{kappa}) reads as \b\label{2.3}
\partial^T\cdot \Psi=x_5[\partial^T.\Phi+4x.\Phi]=2x_{5}x\cdot \Phi, \e
by using (C.3) we can write
\b (Q_{0}-2)\partial^T\cdot \Psi=0.\e  Now, if we act $(Q_0-2)$ on the $\Psi_{\alpha}$  we have
$$(Q_{0}-2)\Psi_{\alpha}=x_5\left[(Q_0-2)\Phi_\alpha+(Q_0-2)x_\alpha x.\Phi)\right]$$
$$ =x_5\left[x_\alpha Q_0x.\Phi-4x_\alpha x.\Phi-2\partial^T_\alpha(x.\Phi)-2x_\alpha x.\Phi \right]$$
 $$ =x_{5}\left[-4x_{\alpha}x\cdot \Phi-2{\partial^T}_{\alpha}(x\cdot \Phi)\right],$$ which has been obtained from the (\ref{kappa}).
Finally the CI spin-$\frac{3}{2}$ equation can be written as \b
(Q_{0}-2)\Psi_{\alpha}+2x_{\alpha}{\partial^T}\cdot \Psi+{\partial^T}_{\alpha}{\partial^T}\cdot \Psi=0.\e

\section{Generalized Legendre functions}

The generalized Legendre
$P^{d+1}_{\lambda}$ introduced in equations (\ref{11.12}) is defined by \cite{Bateman}
\b  P^{d+1}_{\lambda}(Z)=\dfrac{\Gamma(d/2)}{\sqrt{\pi}\Gamma(\dfrac{d-1}{2}}\int^{\pi}_{0}[Z+(Z^2-1)^{1/2}\cos t]^{\lambda}(\sin t)^{(d-2)}dt.\e
 This function is proportional to Gegenbauer function $\textit{C}^{k}_{\lambda}(Z)$ of the first kind as
\begin{eqnarray}
P^{d+1}_{\lambda}(Z)=\dfrac{\Gamma(d-1)\Gamma(\lambda+1)}{\Gamma(\lambda+d-1)}\textit{C}^{(d-1)/2}_{\lambda}(Z)\\ \nonumber =F\left(\lambda+d-1,-\lambda;\dfrac{d}{2};\dfrac{1-Z}{2}\right),
\end{eqnarray}
where \b \textit{C}^{k}_{\lambda}(Z)=\dfrac{\Gamma(\lambda+2k)}{\Gamma(\lambda+1)\Gamma(2k)}F\left(\lambda+2k,,-\lambda,;k+\dfrac{1}{2};\dfrac{1-Z}{2}\right),\e
and $F$ is the hypergeometric function.
The $P^{(d+1)}_{\lambda}(Z)$ and Legendre functions $P^{\mu}_{\nu}(Z)$ are related by
\b  P^{d+1}_{\lambda}(Z)=2^{(d-2)/2}\Gamma(\dfrac{d}{2})(Z^2-1)^{(2-d)/4}P^{(2-d)/2}_{\lambda+(d-2)/2}(Z).\e

\section{Wave equation of spin-3/2 field}

Studying higher-spin fields has a rich history dating back to the early work of Fierz-Pauli (which based on the positivity of energy after first quantization) and  Bargmann-Wigner (which the positivity could be replaced by the requirement  of being an UIR) and many others (see for example \cite{vasi1, vasi2} and references therein).  One approach of studying free higher-spin fields is the analysis of the corresponding relativistic wave equations which could be based on the Bargmann-Wigner scheme.\footnote{We thank a very conscientious referee who suggested this method.}  In this method a manifestly covariant differential equation is associating with a given UIR of the Poincar\'e group and the positive energy solutions of the equations transform according to the corresponding UIR \cite{wigner1}. For example, one can derive
Maxwell's equations via two equations of massless spin-1/2. The solutions of the Bargmann-Wigner equations are actually the fields that transform under the symmetric group as totally symmetric multi-spinors of definite mass and spin. In Anti-de Sitter space, covariant massless fields for half-integer spin have been considered \cite{fronsdal12}. In this appendix, we extend this approach of finding covariant massless field in de Sitter space, by explicitly writing the massless spin-3/2 field equation in terms of vector and spin-1/2 field equations. The mathematical details go as follows.

The action of $S_{\alpha\beta}$ on a tensor field of rank
$l$, $\Psi_{\gamma_1...\gamma_l}(x)$, is given by
       \b S_{\alpha \beta}^{(l)}\Psi_{\gamma_1...\gamma_l}=-i\sum^l_{i=1}
          \left(\eta_{\alpha\gamma_i}
        \Psi_{\gamma_1...(\gamma_i\rightarrow\beta)... \gamma_l}-\eta_{\beta\gamma_i}
          \Psi_{\gamma_1...(\gamma_i\rightarrow \alpha)... \gamma_l}\right).\e
On the other hand, for a half-integer spin field with spin $s=l+\frac{1}{2}$
which is represented by a four component spinor-tensor $\Psi_{
\gamma_{1}...\gamma_{l} }^{i}$ with spinor index $i=1,2,3,4$, one has
$$S_{\alpha\beta}^{(s)}=S_{\alpha\beta}^{(l)}+S_{\alpha\beta}^{(\frac{1}{2})},\qquad
\mbox{with}\qquad
S_{\alpha\beta}^{(\frac{1}{2})}=-\frac{i}{4}\left[\gamma_{\alpha},\gamma_{\beta}\right],$$ where the Dirac gamma matrices are satisfied by (\ref{clifford}) explicitly  given by
$$ \gamma^0=\left( \begin{array}{clcr} I & \;\;0 \\ 0 &-I \\ \end{array} \right)
      ,\gamma^4=\left( \begin{array}{clcr} 0 & I \\ -I &0 \\ \end{array} \right) , $$ \b
   \gamma^1=\left( \begin{array}{clcr} 0 & i\sigma^1 \\ i\sigma^1 &0 \\
    \end{array} \right)
   ,\gamma^2=\left( \begin{array}{clcr} 0 & -i\sigma^2 \\ -i\sigma^2 &0 \\
      \end{array} \right)
   , \gamma^3=\left( \begin{array}{clcr} 0 & i\sigma^3 \\ i\sigma^3 &0 \\
      \end{array} \right),\e
where $\sigma_i$ are Pauli matrices and $I$ is a $2\times2$ unit
matrix. After doing some calculations one can show that for a $l$-rank tensor field
$\Psi_{\gamma_{1}...\gamma_{l}}(x)$, the following relations are hold
 \b Q_l^{(1)}\Psi=Q_0^{(1)}\Psi-2\Sigma_1 \partial x .\Psi+2\Sigma_1 x \partial.
           \Psi+2\Sigma_2 \eta \Psi'-l(l+1)\Psi,\e
where
   \b Q_l^{(1)}=-\frac{1}{2}L_{\alpha \beta}^{(l)}L^{\alpha \beta(l)}
         =-\frac{1}{2}M_{\alpha \beta}M^{\alpha \beta}-\frac{1}{2} S_{\alpha
           \beta}^{(l)}S^{\alpha
            \beta(l)} -M_{\alpha \beta}S^{\alpha \beta(l)},\e

\b M_{\alpha \beta}S^{\alpha \beta(l)}\Psi(x)= 2\Sigma_1
\partial x
       .\Psi-2\Sigma_1 x \partial.\Psi-2l\Psi,\e
       \b\frac{1}{2}S_{\alpha \beta}^{(l)}S^{\alpha \beta(l)}\Psi=l(l+3)\Psi-
          2\Sigma_2 \eta \Psi',\e
where  $Q_0^{(1)}=-\frac{1}{2}M_{\alpha \beta}M^{\alpha \beta}$ and $\Psi'$
is the trace of $\Psi(x)$ and $\Sigma_p$ is the
non-normalized symmetrization operator given by
         \begin{eqnarray}
         \Psi'_{\alpha_1...\alpha_{l-2}}&=&\eta^{\alpha_{l-1}\alpha_l}
          \Psi_{\alpha_1...\alpha_{l-2} \alpha_{l-1}\alpha_l},\nonumber\\
          (\Sigma_p AB)_{\alpha_1...\alpha_l}&=&\sum_{i_1<i_2<...<i_p}
          A_{\alpha_{i_1}\alpha_{i_2}...\alpha_{i_p}}
          B_{\alpha_1...\not\alpha_{i_1}...\not\alpha_{i_2} ...\not\alpha_{i_p}...\alpha_l}.
\end{eqnarray}
In the case of a half-integer spin field, $s=l+\frac{1}{2}$, the
$S_{\alpha\beta}^{(\frac{1}{2})}$ acts only upon the index $i$
$${\cal S}^{(\frac{1}{2})}_{\alpha\beta}{\cal S}^{\alpha\beta(l)}\Psi(x)=l \Psi(x)-\Sigma_1\gamma(\gamma\cdot
\Psi(x)),$$ and the Casimir operator becomes
          \begin{eqnarray}
            Q^{(1)}_s=-\frac{1}{2}\left(M_{\alpha \beta}+S_{\alpha \beta}^{(l)}+
                S_{\alpha\beta}^{(\frac{1}{2})}\right)
      \left(M^{\alpha \beta}+S^{\alpha \beta(l)}+S^{\alpha \beta(\frac{1}{2})}\right)\nonumber\\
    =Q^{(1)}_l-\frac{5}{2}+\frac{i}{2}\gamma_{\alpha}\gamma_{\beta}M^{\alpha
\beta}- S_{\alpha \beta}^{(\frac{1}{2})}S^{\alpha \beta(l)}.
\end{eqnarray}
Therefore one obtains
       \begin{eqnarray}
               Q^{(1)}_s\Psi(x)&=&\left(Q^{(1)}_l-l-\frac{5}{2}+\frac{i}{2}\gamma_{\alpha}\gamma_{\beta}M^{\alpha
       \beta}\right)\Psi(x)+ \Sigma_1 \gamma (\gamma\cdot\Psi(x))\nonumber\\
       &=&\left(-\frac{1}{2}M_{\alpha \beta}M^{\alpha
      \beta}+\frac{i}{2} \gamma_{\alpha}\gamma_{\beta}
       M^{\alpha \beta}-l(l+2)-\frac{5}{2}\right)\Psi(x)\nonumber\\ &-&2\Sigma_1
         \partial x \cdot\Psi(x)+2\Sigma_1 x \partial\cdot\Psi(x)+2\Sigma_2 \eta
       \Psi'(x)+\Sigma_1 \gamma (\gamma\cdot \Psi(x)).
        \end{eqnarray}
Therefore, any higher-spinor field, $s=l+\frac{1}{2}$, can indeed be decomposed into a tensor and spinor field. In our case in this paper it was shown that $s=\frac{3}{2}$ field can be written in terms of a spinor field and a polarization vector with total four degrees of freedom in de Sitter space.

\end{document}